\documentclass[a4paper,10pt,twocolumn,superscriptaddress,aps,pre]{revtex4-1}
\usepackage[latin1]{inputenc}
\usepackage{amsmath}
\usepackage{amsfonts}
\usepackage{amssymb}
\usepackage{graphicx}
\usepackage{color}
\usepackage{soul}
\usepackage[normalem]{ulem}

\graphicspath{{figures/}}

\begin{document}
\title{Sequential disruption of the shortest path in critical percolation}
\date{\today}

\author{Oliver Gschwend}
\affiliation{ETH Z\"urich, Computational Physics for Engineering Materials, Institute for Building Materials, Wolfgang-Pauli-Str. 27, HIT, CH-8093 Z\"urich (Switzerland)}
\author{Hans J. Herrmann}

\affiliation{Departamento de F\'isica, Universidade do Cear\'a, 60451-970 Fortaleza (Brazil)}
\affiliation{ESPCI, CNRS UMR 7636 - Laboratoire PMMH, 75005 Paris (France)}
\begin{abstract}
\noindent We investigate the effect of sequentially disrupting the shortest path of percolation clusters at criticality by comparing it with the shortest alternative path. We measure the difference in length and the enclosed area between the two paths. The sequential approach allows to study spatial correlations. We find the lengths of the segments of successively constant differences in length to be uncorrelated. Simultaneously, we study the distance between red bonds. We find the probability distributions for the enclosed areas $A$, the differences in length $\Delta l$, and the lengths between the red bonds $l_r$ to follow power law distributions. Using maximum likelihood estimation and extrapolation we find the exponents $\beta = 1.38 \pm 0.03$ for $\Delta l$, $\alpha = 1.186 \pm 0.008$ for $A$ and $\delta = 1.64 \pm 0.025$ for the distribution of $l_r$.
\end{abstract}

\maketitle

\section{Introduction}
\label{Introduction}
The optimum path through a random energy landscape has been studied exhaustively in the past \cite{Porto1997, Porto1999, Cieplak1994, Cieplak1996}. In particular also the blocking of sites along the optimum path and the resulting second best path has been considered \cite{oliveira2011}. Percolation clusters at criticality are realizations of fractal uncorrelated random landscapes and in the present paper we will study blocking the shortest path. We consider site percolation on a square lattice at the critical threshold. Assuming the model describes a traffic situation \cite{Li2015} or a forest fire \cite{mackay1984}, it is important to know how far the bypass and the surrounded area will be in case of a blockade in the shortest path. 
A similar investigation to ours was performed before \cite{Hillebrand2018} but on a directed bond percolation grid and without the sequential approach which allows to obtain the spatial correlations along the path. Nevertheless, due to strong analogies between the two models \cite{christensen2005}, we expect to obtain similar results.

\section{Model}

\begin{figure}[ht]
	\begin{center}
		\includegraphics[scale=0.35]{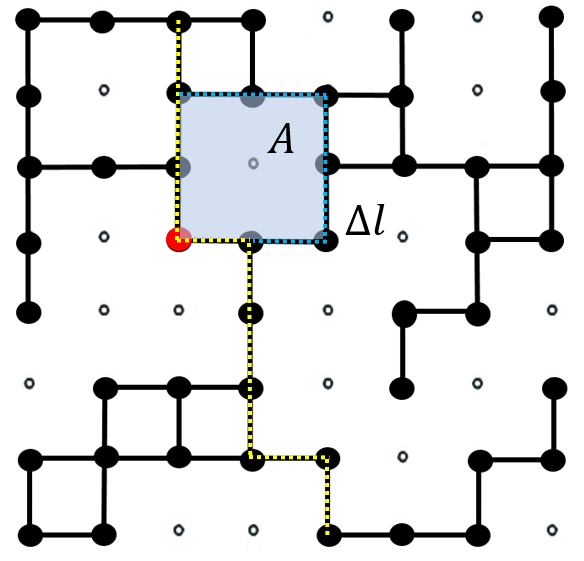}
	\end{center}
	\caption{A percolating cluster with the shortest path (yellow dashed),a site from the shortest path blocked (red), alternative second shortest path (blue dashed) and the smallest enclosed area $A$  (transparent blue). $\Delta l$ is the difference in length between the two paths, here $\Delta l = 2$.}
	\label{fig:1}
\end{figure}

We simulate site percolation on a two dimensional square lattice at the percolation threshold $p_c = 0.592746$ \cite{Newman2000}. In the horizontal direction (left, right) we choose periodic boundary conditions to reduce finite size effects. In vertical direction the upper and the lower boundaries are open. Using the burning algorithm \cite{Herrmann1984}, we first extract the shortest path $l$ or 'chemical distance' \cite{Havlin1984}, between the upper and the lower border of the lattice, which is known to be fractal at $p=p_{c}$ \cite{Grassberger1983, Pike1981, Herrmann1984, Grassberger1985,Herrmann1988}. In our case we consider only systems in which two completely separated shortest paths do not exist. In case, the shortest path is not unique, we choose randomly one. Next, we walk along the shortest path, sequentially blocking each time one site on the path and then run the burning algorithm again to find the shortest alternative path. If there are several second shortest paths of equal length we choose the one enclosing the smallest area with the original one. After each blocking event, the blocked site will be unblocked again. An example is shown in Fig.~\ref{fig:1}, where the original shortest path (yellow line with black dots) is blocked (red site) and the shortest alternative path is found (blue line with black dots). The enclosed area is filled in light blue and the difference in length is obtained by subtracting the lengths of the two paths. 
In order to obtain good statistics we perform at least 1.4 million blocking events per lattice size, which ranges from $L=200$ up to $L=15000$. We choose the functional form of the probability density function by comparing the log-likelihood ratios from; lognormal, exponential, power-law, and exponentially truncated power-law. According to Ref.\cite{White2008}, we use maximum likelihood estimation (MLE) for discrete exponentially truncated power-laws based on the method used in Ref. \cite{alstott2014}. The estimated exponents for different lattice sizes~$L$ show algebraic convergence to some limit value $\gamma_0$ as $L\rightarrow\infty$.  By fitting the estimated exponents~$\gamma(L)$ against $\gamma(L) = \gamma_0 + \gamma_s L^{\gamma_e}$ using non-linear least squares, we extrapolate $\gamma_0$. Since the probability distribution in section~\ref{difference in length}, Fig. \ref{fig:4} follows a power-law which is flattened in the beginning, we can not use the method from Ref.~\cite{alstott2014}. Instead we use a least squares fit applied on 12 normalized log-bins using the points number 3 to 11. 

\section{distances between successive red bonds}
\label{lengths between red bonds}

We start by extracting the shortest path and plotting the average length of the shortest paths $\langle l \rangle$ against the lattice size. We obtain the fractal dimension of the shortest path  $\langle l \rangle \sim L^{d_{min}}$ with $d_{min} = 1.123 \pm 0.015$, which is within the error-bars in good agreement with Ref. \cite{Zhou2012}. 
 
\begin{figure}[ht]
    \includegraphics[scale=0.42]{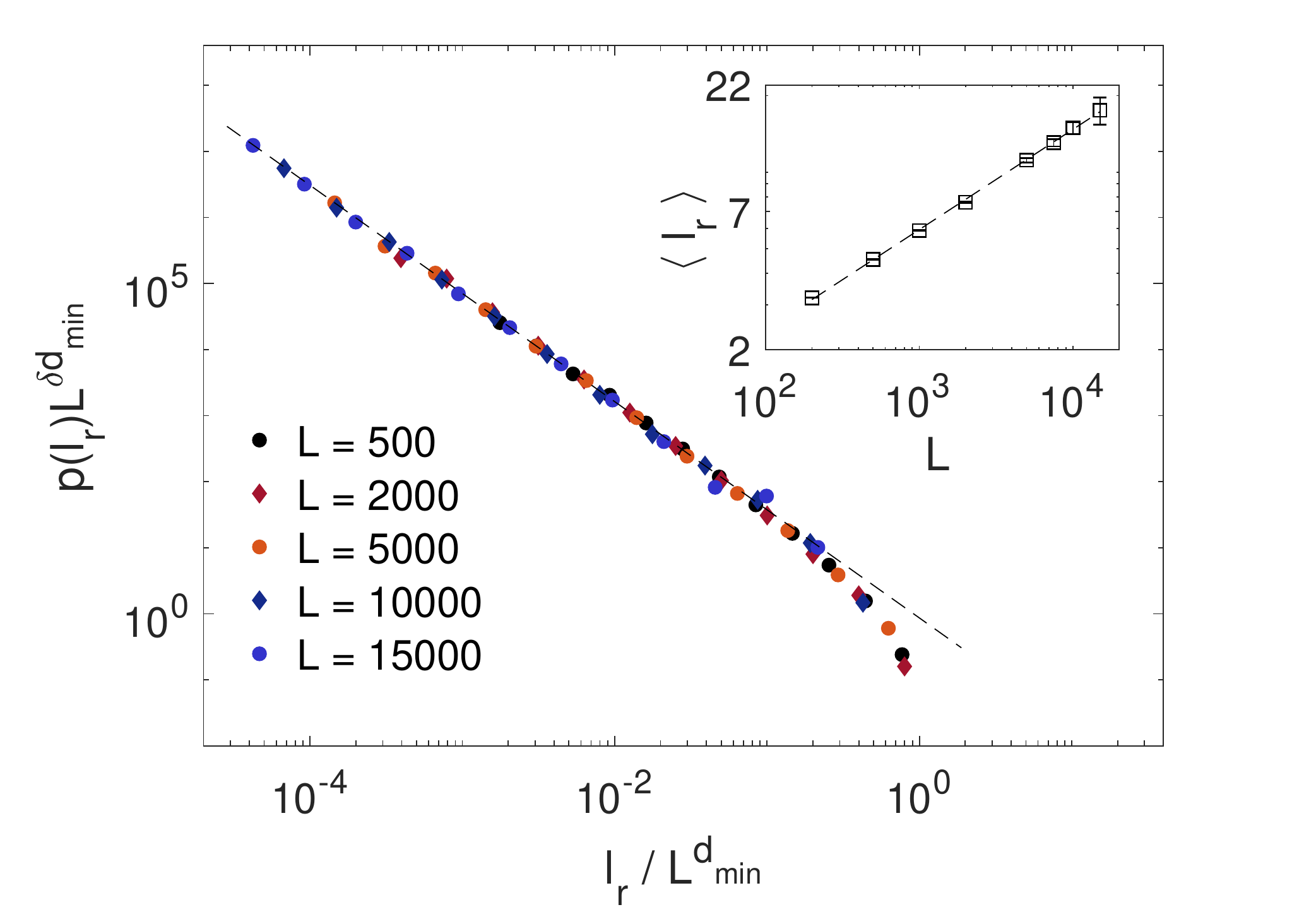}
	\caption{Data collapse for the probability distribution function for the distances between red bonds $p(l_{r})$ for different lattice sizes $L$. The dashed line represents a power law fit with exponent $\delta = 1.64 \pm 0.025$. The inset shows the average length between red bonds $\langle l_{r} \rangle$ for different lattice sizes $L$. The dashed line is a power law fit with exponent $0.39 \pm 0.02$ giving the lattice size dependency of $\langle l_{r} \rangle$.}
	\label{fig:2}
\end{figure}

Next we start blocking the shortest path at the topmost site on the upper border and wander along the shortest path until the lower border is reached. This process of blocking all sites in a successive order we call 'sequential disruption'. Each time we block a site, we look for the shortest alternative path through the lattice. In case where no alternative path through the lattice exists, we are blocking a cutting or red bond \cite{Pike1981}. A red bond is defined as a site belonging to the shortest path, which when blocked disrupts the whole connection from the top to the bottom. Supposing that the locations of red bonds along the shortest path are uncorrelated \cite{Herrmann1984}, we propose that the average length between red bonds $\langle l_{r} \rangle$ scales as the ratio of the shortest path length $\langle l \rangle$ and the number of red bonds $\langle n_{r} \rangle$

\begin{equation}
	\langle l_{r} \rangle \sim \frac{\langle l \rangle}{\langle n_{r} \rangle} = L^{d_{min}-\frac{1}{\nu}}
    \label{eq:one}
\end{equation}

\noindent because the number of red bonds scales with $\langle n_{r} \rangle \sim L^{\frac{1}{\nu}}$ \cite{Coniglio1982}, where $\nu = 4/3$ is the exponent of the divergence of the correlation length \cite{Stauffer2014} and the shortest path scales with $\langle l \rangle \sim L^{d_{min}}$, with $d_{min} = 1.13077(2)$ \cite{Zhou2012}. As shown in the inset of Fig.~\ref{fig:2}, we find that  $\langle l_{r} \rangle$ scales with $L$ as $L^{0.39 \pm 0.02}$, which is within error bars in agreement with Eq.~(\ref{eq:one}).  

We also study their empirical probability distribution. We find the probability distribution of $l_{r}$ to follow an exponentially truncated power law. We propose the following scaling ansatz:

\begin{equation}
	p(l_{r},L) = L^{-\delta d_{min}} F(\frac{l_{r}}{L^{d_{min}}})
	\label{eq:two}
\end{equation}

We divide the argument of the scaling function $F$ by $L^{d_{min}}$ motivated by the fact that $l_{r}$ is naturally limited by the shortest path length. Using that the scaling function  $F(x) \sim x^{-\delta}$ for $x\rightarrow\infty$, we have to multiply the scaling function with $L^{-\delta d_{min}}$. We confirm our scaling ansatz with the good data collapse shown in Fig.~\ref{fig:2} with the exponent $\delta = 1.64 \pm 0.025$. The truncation is due to the finite size of the systems and is expected to vanish for $L\rightarrow\infty$.

\section{difference in length between shortest paths}
\label{difference in length}
\begin{figure}[ht]
    \includegraphics[scale=0.42]{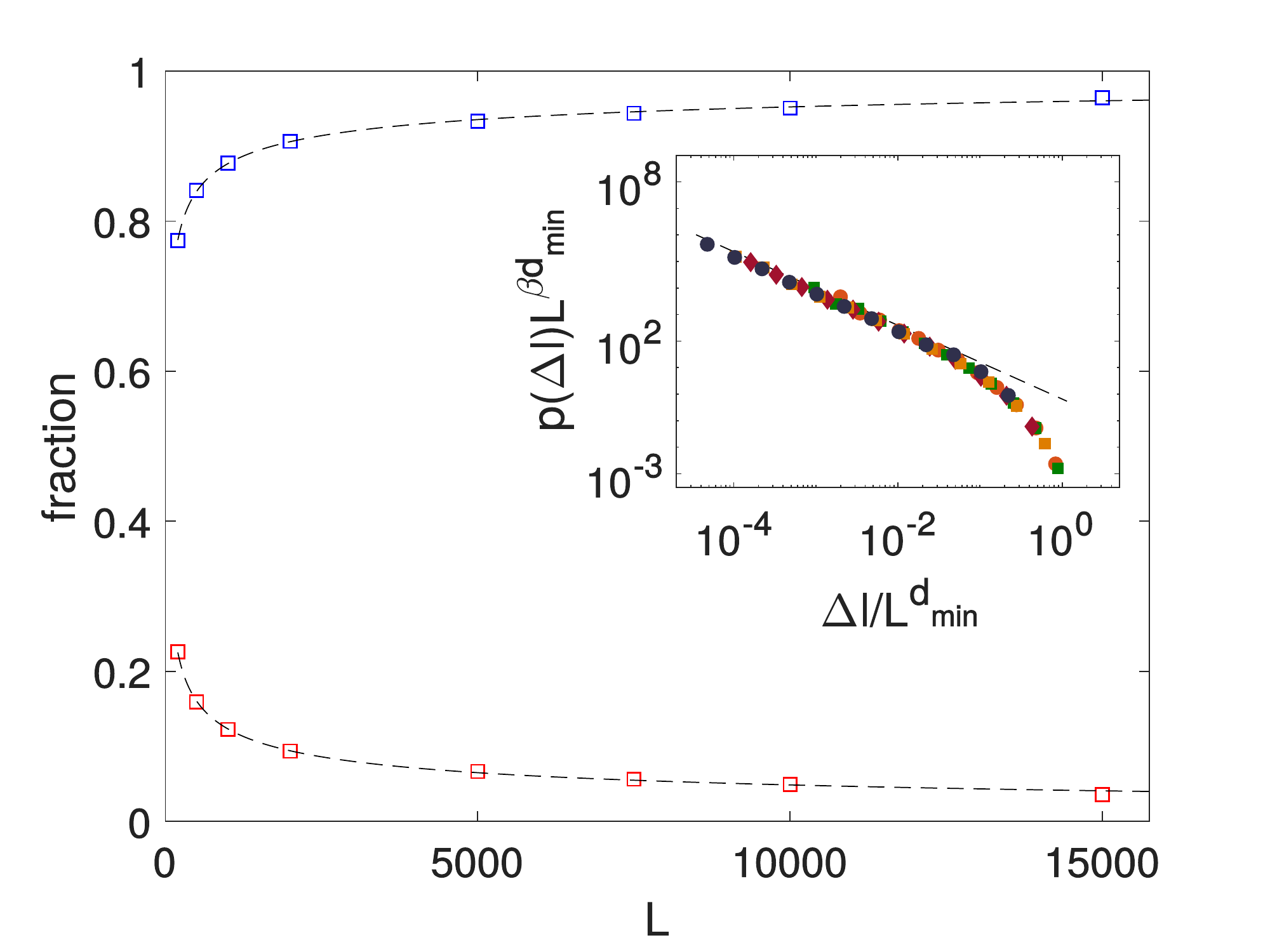}
	\caption{The fraction of convergent and divergent alternative paths. The dashed line represents the algebraic fit. The inset shows the data collapse for the probability distributions for the difference in length $\Delta l$ for different lattice sizes, ranging from $L = 500$ up to $L = 15000$. The dashed line depicts a power law fit with exponent $\beta = 1.38 \pm 0.03 $.}
	\label{fig:3}
\end{figure}

In this section we discuss the results concerning the difference in length between the shortest path and the next shortest alternative after removing one site of the shortest path. We observe two cases of alternative paths. They can either have the same (convergent) or different (divergent) starting and end points. In the latter case the two paths together form an open fork towards the upper or lower border of the system. As shown in Fig.~\ref{fig:3}, we measure the fraction of each case. Using an algebraic fit of the form $f=f_{0}+p_{1}f(L)^{p_{2}}$, where $f_{0}$ is the extrapolated value of the fractions for $L\rightarrow\infty$, we find that the fraction of divergent cases tends to $-0.01 \pm 0.03$, which suggests that the divergent case might be just a finite size effect and disappears in the infinite limit. It might also be that the fraction of divergent paths does go to a very small number in the thermodynamic limit \cite{cardy1998number,grassberger2002}. Consequently, we include only the convergent case in further studies.

 We find the probability distribution of $\Delta l$ to follow an exponentially truncated power-law. We propose the scaling ansatz of the form: 
 
 \begin{equation}
	p(\Delta l,L) = L^{-\beta d_{min}} G(\frac{\Delta l}{L^{d_{min}}})
	\label{eq:three}
\end{equation}

The division of the argument of the scaling function $G$ by $L^{d_{min}}$ is motivated by the fact that the fractal dimension of shortest distance follows $\langle l \rangle \sim r^{d_{min}}$ \cite{Zhou2012}. We find the best model to be an exponentially truncated power law. Indeed, we observe a convincing data collapse shown in the inset of Fig.~\ref{fig:3} and hence confirm our proposed scaling ansatz with an estimated exponent of $\beta = 1.38 \pm 0.03$. This value is consistent with the one found on a randomly directed square lattice in Ref.~\cite{Hillebrand2018}.

\begin{figure}[ht]
    \includegraphics[scale=0.42]{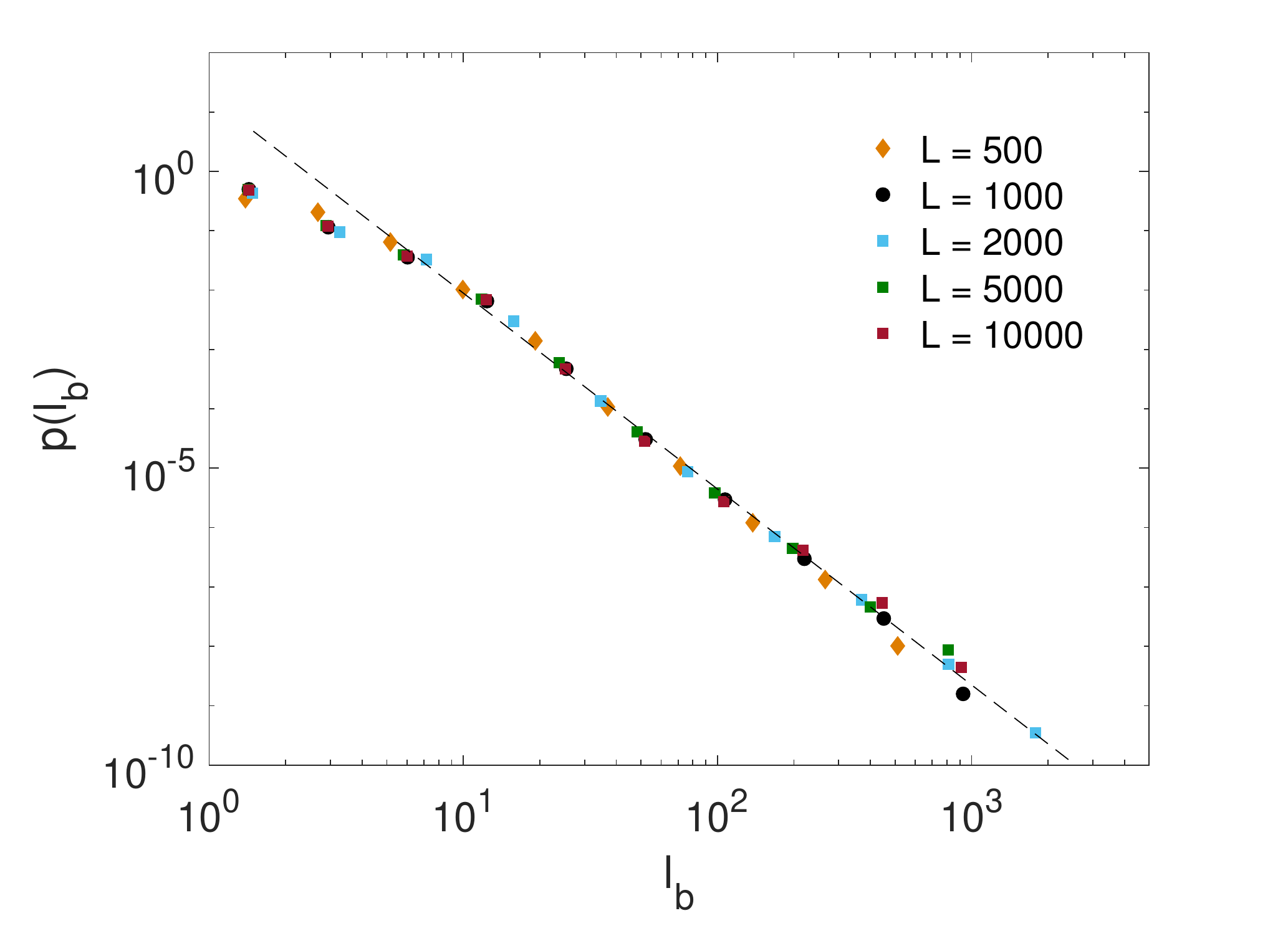}
	\caption{Probability distribution of the segment lengths $p(l_{b})$ for different lattice sizes. The dashed line indicates a power law fit with exponent $\epsilon = 3.3 \pm 0.1$}
	\label{fig:4}
\end{figure}

We now consider the spatial correlation of the detours. Therefore, we walk along the shortest path and count over how many blocked sites the difference in length $\Delta l$ between the shortest path and the next shortest path stays the same. Successive sites which show the same difference in length belong to the same "segment". As an example consider in Fig.~\ref{fig:1}, instead of the red indicated site, the previous one being blocked. Both, the alternative path and the difference in length stay the same and hence the two successive sites belong to the same segment. We measure the lengths of these segments $l_b$ and obtain them in a spatially ordered sequence. By calculating the sample autocorrelation function of the spatial sequence of segment lengths, we find that the series is uncorrelated. Meaning, that the length of one segment is independent of the length of the previous one. We present the distribution of the segment-lengths in Fig.~\ref{fig:4}.

The probability distribution of $l_{b}$ follows a power law with an estimated exponent of $\epsilon = 3.3 \pm 0.1$. Hence, the probability of finding longer constant segments decreases rapidly.

\section{enclosed Area}
\label{enclosed Areaa}
Next, we study the enclosed area between the shortest path and its next shortest alternative. We consider only the well defined areas surrounded by a closed loop formed by the two paths. Enclosed areas are apparently compact objects and hence we divide the argument of the proposed scaling function by $L^2$ to collapse the data. We find that the probability distribution function $H$ follows an exponentially truncated power law. Consequently, we propose a scaling law of the form:

\begin{equation}
	p(A,L) = L^{-2 \alpha} H(\frac{A}{L^2})
	\label{eq:four}
\end{equation}

\begin{figure}[ht]
    \includegraphics[scale=0.42]{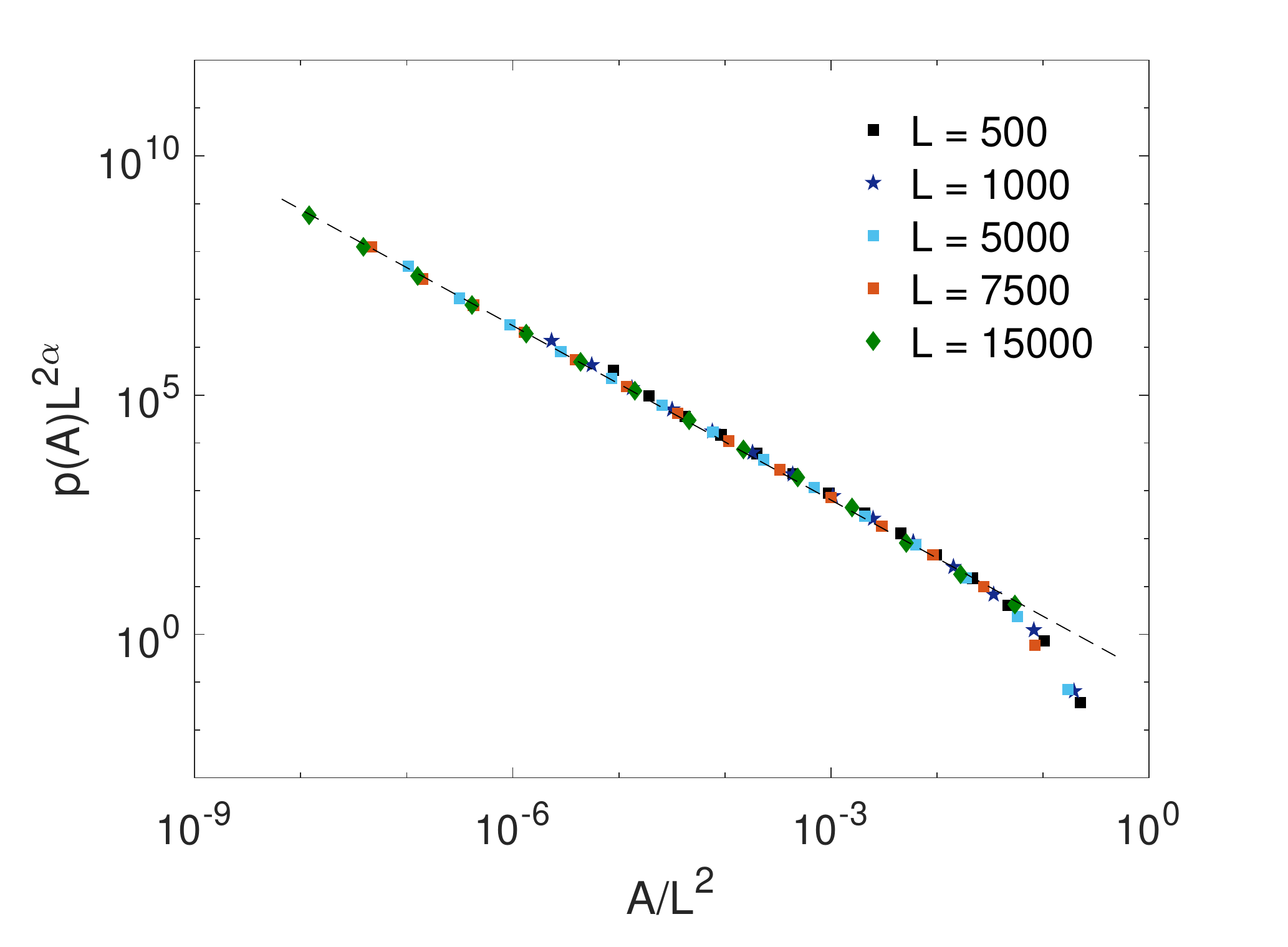}
	\caption{Data collapse for the probability distribution functions of the areas enclosed by the shortest path and its next shortest alternative $p(A)$ for different lattice sizes $L$. The dashed line indicates a power law fit with exponent $\alpha = 1.186 \pm 0.008$ }
	\label{fig:5}
\end{figure}

We obtain an excellent data collapse shown in Fig.~\ref{fig:5} which verifies our proposed scaling law with an estimated exponent of $\alpha = 1.186 \pm 0.008$. Our result is consistent with previous findings on a randomly directed square lattice \cite{Hillebrand2018} where an exponent of $1.189 \pm 0.001$ was found and on an artificial landscape where the area between watersheds was investigated \cite{Fehr2011} where an exponent of $1.16 \pm 0.03$ was reported.

\section{Conclusion}
We studied the effect of sequentially blocking each site along the shortest path of a critical two dimensional site percolation cluster by considering the shortest alternative path. We derived the average distance between the red bonds theoretically and confirmed it with numerical results. We found the lengths between the red bonds to be power law distributed and presented a scaling law with an exponent $\delta = 1.64 \pm 0.025$.
The comparison between the two paths provides us with the difference in their length and the enclosed area between them. We found the size of the smallest enclosed areas $A$ and the differences in length $\Delta l$ to be power law distributed with exponents $\alpha = 1.186 \pm 0.008$ and $\beta = 1.38 \pm 0.03$ using MLE and verified our proposed scaling law by collapsing the data. Further, we counted over how many successive blocked sites $\Delta l$ stays constant and called this quantity $l_{b}$. We found the series of spatially ordered segment lengths to be uncorrelated by calculating the sample autocorrelation function. We presented the distribution of $l_{b}$, which follows a power law with an estimated exponent $\epsilon = 3.3 \pm 0.1$, using least squares on the log-binned histogram.  

Since percolation describes many natural processes like forest fires \cite{mackay1984}, electrical breakdown \cite{Niemeyer1984} or traffic conditions \cite{Li2015}, it would be of interest to link the blocking approach to real systems. Also other shortest alternative paths enclosing not only the smallest area, but rather all possible areas could be investigated.

\acknowledgments
HJH thanks Funcap and CAPES.


%

\end{document}